\documentclass[aps,prd,showpacs,floatfix,preprintnumbers,nofootinbib,
amssymb,amsmath]{revtex4}
\usepackage{graphicx,bm,color}

\begin{document}

\title{Fluctuation of strongly interacting matter in Polyakov Nambu
Jona-Lasinio model in finite volume }

\author{Abhijit Bhattacharyya}
\email{abphy@caluniv.ac.in}
\affiliation{Department of Physics, University of Calcutta,
92, A. P. C. Road, Kolkata - 700009, INDIA}
\author{Rajarshi Ray}
\email{rajarshi@jcbose.ac.in}
\affiliation{Department of Physics and Centre for Astroparticle Physics
\& Space Science, Bose Institute,
EN-80, Sector V, Bidhan Nagar, Kolkata - 700091, INDIA}
\author{Subrata Sur}
\email{ssur.phys@gmail.com}
\affiliation{Department of Physics, Panihati Mahavidyalaya, Barasat
Road, Sodepur, Kolkata - 700110, INDIA}

\begin {abstract}
We estimate the susceptibilities of conserved charges for two flavor
strongly interacting matter with varying system sizes, using the
Polyakov loop enhanced Nambu--Jona-Lasinio model. The susceptibilities
for vanishing baryon densities are found to show a scaling with the
system volume in the hadronic as well as partonic phase. This scaling
breaks down for a temperature range of about 30-50 MeV around the
crossover region. Simultaneous measurements of the various
susceptibilities may, thus, indicate how close to the crossover region
the freeze-out occurs for the fireball created in heavy-ion collision
experiments.
\end{abstract}
\pacs{12.38.Aw, 12.38.Mh, 12.39.-x}

\maketitle

Strongly interacting matter under extreme conditions of temperature and
density is expected to show a rich phase structure. In the early Universe, a
few microsecond after the Big Bang, when the temperature was extremely
high, exotic states namely quarks and gluons may have been prevalent~\cite{kolb}. 
On the other hand inside the core of a compact star, where
the baryon matter density is extremely high various exotic phases like
color superconductor, color superfluid etc. may be present~\cite{rajagopal}. 
Experiments with heavy ions colliding with each other
or with a target, at the facilities at CERN(France/Switzerland), BNL(USA)
and GSI(Germany), are continuing the search for such exotic states of
matter in the laboratory.

The matter formed due to the energy deposition of the colliding
particles obviously has a finite volume. It is therefore imperative to
have a clear understanding of the finite size effects to fully
contemplate the thermodynamic phases that may be created in the
experiments. These effects depend on the size of the colliding nuclei,
the center of mass energy ($\sqrt s$) and the centrality of collisions. 
There have been a large number of efforts to estimate the system size
at freeze-out for different $\sqrt s$ and different centralities. A
study using the measurement of HBT radii~\cite{ceres} indicates that the
freeze out volume increases as the $\sqrt s$ increases. In the same work
the freeze out volume has been estimated to be  $2000~fm^3$ to
$3000~fm^3$. On the other hand in Ref.~\cite{graf} the volume of
homogeneity has been calculated using UrQMD model~\cite{urqmd} and
compared with the experimentally available results. The $\sqrt s$
considered was in the range of 62.4 GeV to 2760 GeV for lead-lead
collisions at different centralities. The system volume has been found
to vary from $50~fm^3$ to  $250~fm^3$. The effect of colliding particles
and $\sqrt s$ has been further analyzed by the ALICE collaboration in~\cite{alice-fs}. 
Given that these are the volumes at the time of
freeze-out, one may expect an even smaller system size at the initial
equilibration time~\cite{initi-fs1,initi-fs2}.

The importance of finite size effects in the thermodynamics of strong
interaction may be brought forward with the help of finite size scaling 
analysis~\cite{fss}. In the context of heavy ion collisions such a
possible analysis has been discussed in the literature
(see e.g.~\cite{fss1,fss2,fss3}).

Other theoretical studies of finite volume effects have been performed
in various contexts. In Ref.~\cite{elze} the effective degrees of
freedom have been found to be reduced due to finite volume using a
non-interacting bag model. The effect of finite volume has been studied
also with a  two model equation of state and it has been found that 
the critical temperature looses its sharpness~\cite{spieles}. A few
first principle study of pure gluon theory on space-time lattices were
performed, showing the possibility of significant finite size
effects~\cite{fslat1,fslat2}. The meson properties show a significant
volume dependence  as found in Ref.~\cite{fischer1,fischer2}. In the
context of chiral perturbation theory the implications of finite system
size have been discussed~\cite{lusher1,gasser1}. There are also studies
with four-fermi type interactions, like the Nambu$-$Jona-Lasinio (NJL)~\cite{Nambu} 
models~\cite{kiriyama,fss1,shao}, linear sigma models~\cite{fss2,braun1,braun2} 
and Gross-Neveu models~\cite{khanna}. 
While in Ref.~\cite{braun1} the scaling behavior of chiral phase
transition for finite and infinite volumes has been studied, the
character of phase diagram has been studied in
Ref.~\cite{fss1,fss2,braun2,khanna}. In refs.~\cite{kiriyama} and~\cite{shao} 
the authors have studied the chiral properties as a
function of the radius of a finite droplet of quark matter. The
stability of such a droplet in the context of strangelet formation
within the NJL model has been addressed in Ref.~\cite{yasui}.
Size dependent effects of di-fermion states within 2-dimensional NJL
model has been studied in Ref.~\cite{abreu1} and that of magnetic field
is discussed in Ref.~\cite{abreu2}. Recently in a 1+1 dimensional NJL
model the induction of charged pion condensation phenomenon in dense
baryonic matter due to finite volume effects have been studied in~\cite{ebert}. 
Recently some of us have studied the thermodynamic
properties of strongly interacting matter in a finite volume using
Polyakov-Nambu-Jona--Lasinio (PNJL) model~\cite{ab}. It has been 
shown there that the critical temperature for the cross-over transition
at zero baryon density decreases as the volume decreases. Furthermore 
at low volume the critical end point is pushed towards the higher $\mu$ 
and lower $T$ domain. At $R=2fm$, it was found that the critical end
point (CEP) vanishes and the whole phase diagram becomes a cross-over.
The possible chiral symmetry restoration in a color confined state
has also been discussed.

Though various thermodynamic properties have been studied to some
extent in finite size systems, not much has been done to estimate
the fluctuations occurring in finite volumes. On the lattice, the
Polyakov loop susceptibility has been calculated for a finite
volume~\cite{berg}. A similar work has been done in the PNJL model
using Monte Carlo simulation~\cite{cristoforetti} and also in the
quark-meson model using a renormalisation group approach~\cite{tripolt}.
Fluctuations of conserved quantum numbers are related to the respective
susceptibilities via the fluctuation-dissipation theorem. For a
2-flavor strongly interacting system one has the quark number
susceptibility (QNS) and isospin number susceptibility (INS) etc. These
fluctuations are sensitive indicators of the transition from hadronic
matter to partonic state. Also the existence of the CEP may be
signalled by the diverging behavior of fluctuations.
Here we report our calculations of quark and isospin number
susceptibilities of strongly interacting matter using PNJL model up
to sixth order. The report is organized as follows. First we give a
very brief description of the PNJL model and the necessary methodology.
Thereafter we present the results for the various susceptibilities.
Finally we summarize and conclude.

The PNJL model used here is based on a series of works
~\cite{Nambu,ab,klev,hat1,fuku,ratti,MAS3_2006,
ray1,ray2,MAS_12,MAS_14,aminul1,MAS1_2006,MAS2_2006,Muller}. For some
recent progress in this model see e.g.~\cite{shear,vphimu,isospin,
aminul1,quarkyonic,shear1,charge,avg_phase_pnjl,CP_pnjl,color_pnjl1,
color_pnjl2}. For a detailed overview see e.g.~\cite{ray3} and 
references therein. The PNJL model for 2 flavors is
described by the Lagrangian,  
\begin {align}
   {\cal L} &= {\sum_{f=u,d}}{\bar\psi_f}\gamma_\mu iD^\mu
             {\psi_f}-\sum_f m_{f}{\bar\psi_f}{\psi_f}
              +\sum_f \mu_f \gamma_0{\bar \psi_f}{\psi_f}\nonumber\\
     &+{\frac {g_S} {2}} {\sum_{a=1,2,3}}[({\bar\psi} \tau^a {\psi})^2+
            ({\bar\psi} i\gamma_5\tau^a {\psi})^2] 
        -{\cal {U^\prime}}(\Phi[A],\bar \Phi[A],T)
\label{lag1}
\end {align}
where Polyakov loop potential ${{\cal {U^\prime}}
(\Phi[A],\bar \Phi[A],T)}$ can be expressed as,
\begin{equation}
\frac {{\cal {U^\prime}}(\Phi[A],\bar \Phi[A],T)} {T^4}= 
\frac  {{\cal U}(\Phi[A],\bar \Phi[A],T)}{ {T^4}}-
                                     \kappa \ln(J[\Phi,{\bar \Phi}])
\label {uprime}
\end{equation}
Here ${\cal U}(\Phi,{\bar \Phi},T)$ is a Landau-Ginsburg type potential
as given in Ref.~\cite{ratti},
\begin{equation}
\frac  {{\cal U}(\Phi, \bar \Phi, T)}{  {T^4}}=-\frac {{b_2}(T)}{ 2}
                 {\bar \Phi}\Phi-\frac {b_3}{ 6}(\Phi^3 + \bar \Phi^3)
                 +\frac {b_4}{  4}{(\bar\Phi \Phi)}^2,
\end{equation}
where
\begin {eqnarray}
     {b_2}(T)=a_0+{a_1}(\frac { {T_0}}{ T})+{a_2}(\frac {{T_0}}{ T})^2+
              {a_3}(\frac {{T_0}}{T})^3,
\end {eqnarray}
$b_3$ and $b_4$ being constants. The second term in Eqn.(\ref{uprime})
is the Vandermonde term.
\begin {equation}
J[\Phi, {\bar \Phi}]=(27/24{\pi^2})\left[1-6\Phi {\bar \Phi}+\nonumber\\
4(\Phi^3+{\bar \Phi}^3)-3{(\Phi {\bar \Phi})}^2\right]
\end{equation}
The parameters $a_i$, $b_i$ were fitted from Lattice results of pure
gauge theory. The set of values chosen here are,
\begin{center}
$a_0=6.75$, $a_1=-1.95$, $a_2=2.625$, $a_3=-7.44$, $b_3=0.75$,
  $b_4=7.5$, $T_0=190 MeV$, $\kappa =0.2$  
\end{center}

To incorporate the effect of finite volume we use a non-zero low
momentum cut-off $p_{min}=\pi/R=\lambda$ where $R$ is the lateral
size of a cubic volume $V=R^3$. In principle one should sum over
discrete momentum values but for simplification we integrate over
continuous values of momentum. Also we neglect surface and curvature
effects. Other parameters of the model were not modified.

We note here that in the NJL model discussions of using a lower
momentum cut-off exists in the literature (see~\cite{ebert1,blaschke}
and references therein). The motivation of introducing this IR cut-off
there has been to mimic confining effects of strong interaction which
helps to remove spurious poles in the quark loop diagrams so that
unphysical decay of hadrons to quarks do not take place. Since in the
PNJL model such unphysical decays are restricted due to the vanishing
of the Polyakov loop for low temperatures~\cite{ratti,blaschke1} no IR
cut-off is necessary in the PNJL model. However, for 2 flavour PNJL model, 
the unphysical decay 
does not completely vanish as pointed out in Ref.~\cite{prd75} where a 
very small but non-zero sigma meson decay is observed at low temperatures. 
On the other hand, for 2+1 flavour the $\sigma$-meson becomes a true bound state 
for small temperatures~\cite{prd79}. 
\begin{figure}[htb]
\includegraphics[scale=0.45]{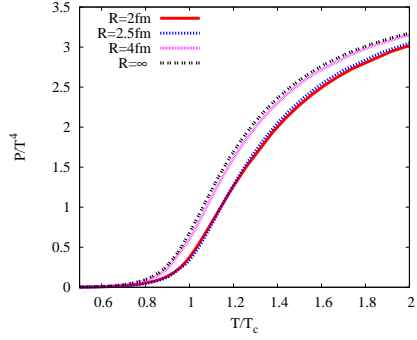}
\caption{(Color online) 
Variation of pressure with temperature for different system sizes.}
\label {pressure}
\end{figure}

\begin{figure}[htb]
\includegraphics[scale=0.45]{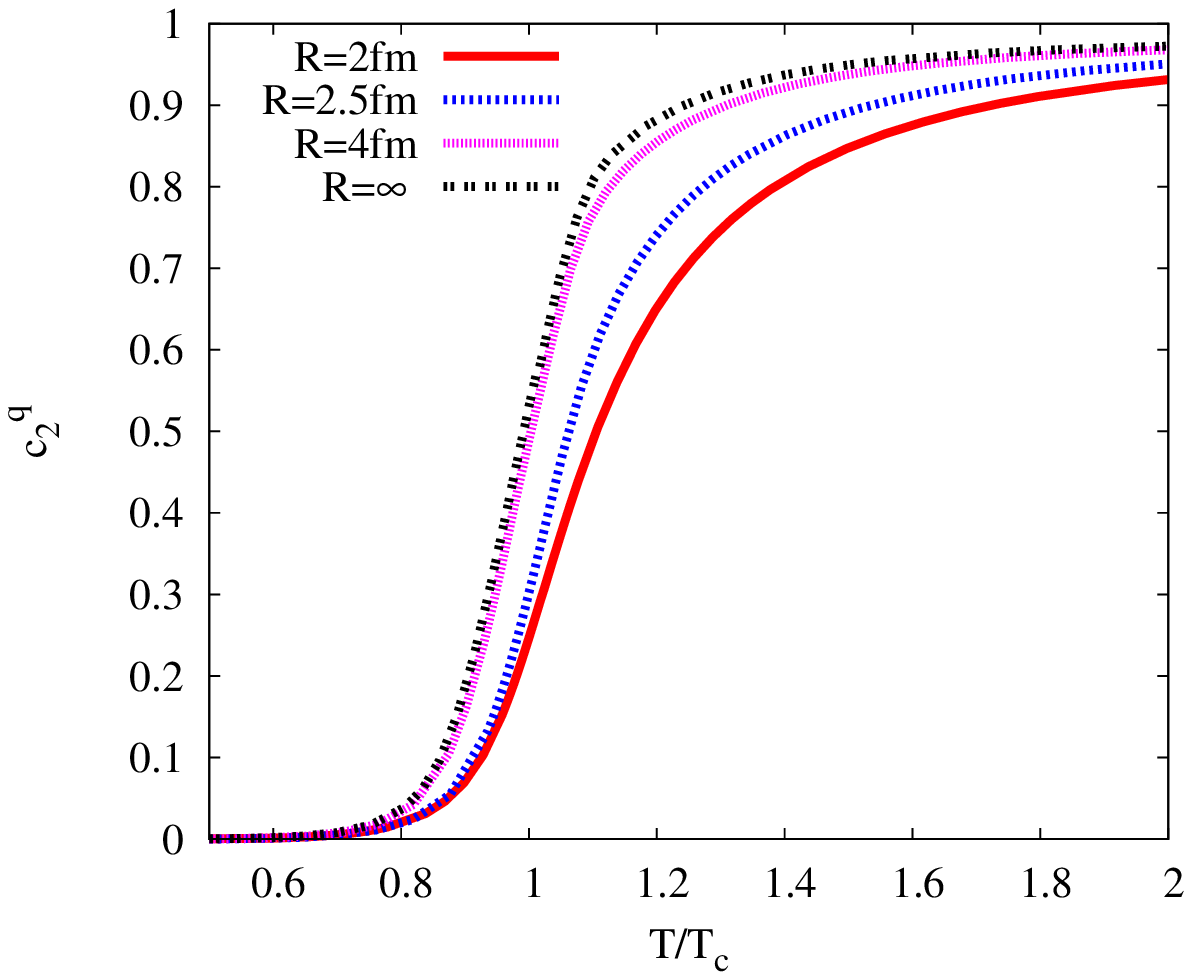}
\includegraphics[scale=0.45]{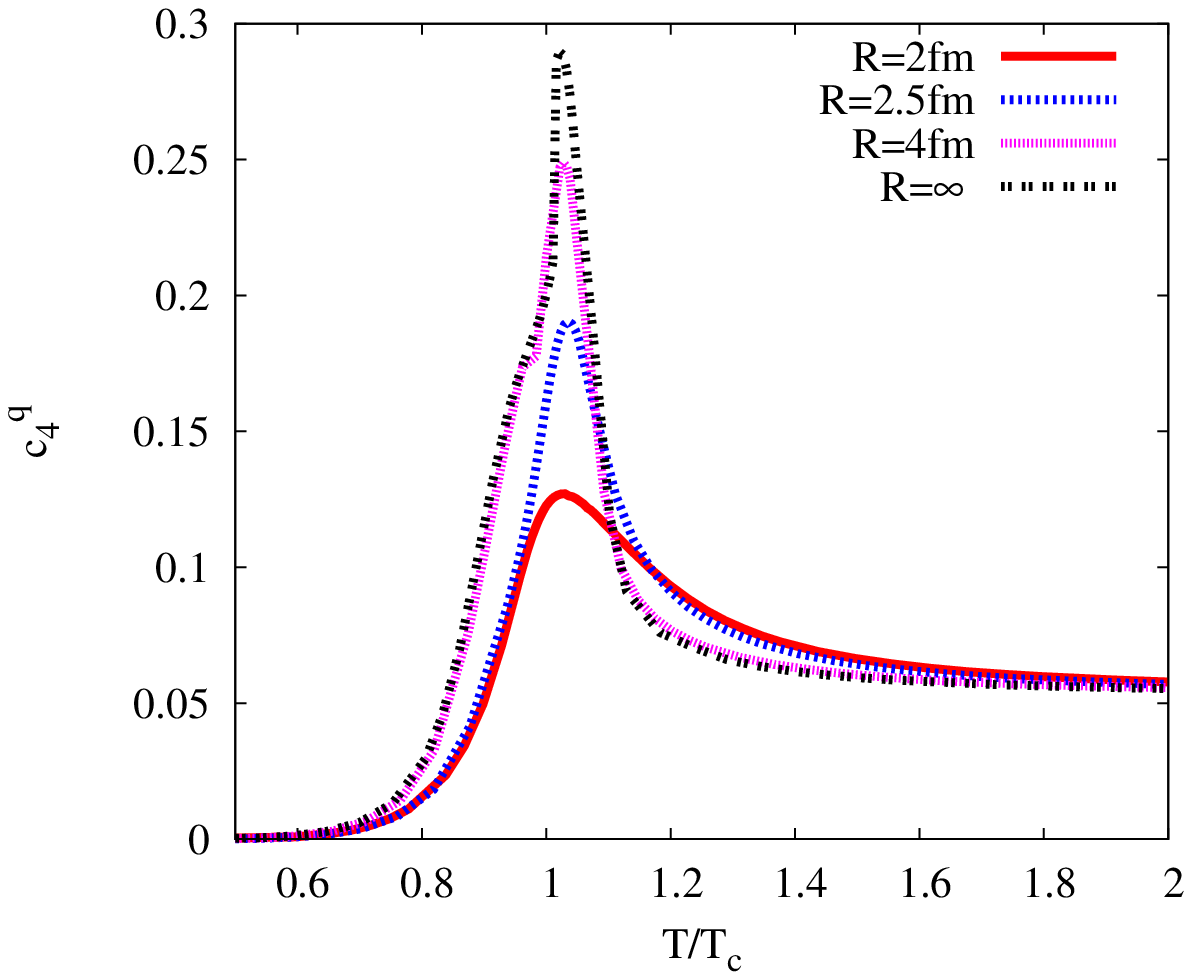}
\includegraphics[scale=0.45]{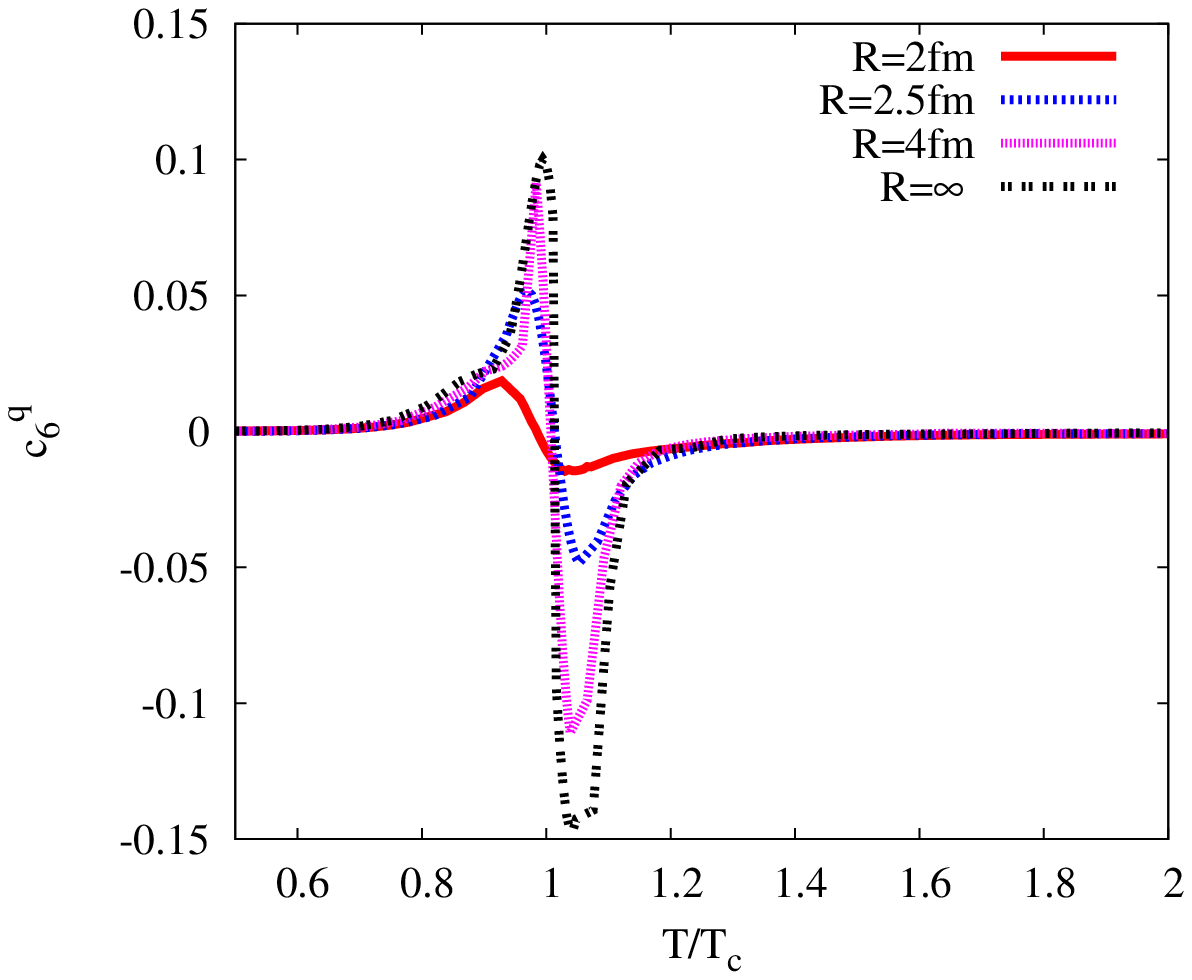}
\caption{(Color online) 
Variation of quark number susceptibilities with temperature for different
system sizes.}
\label {quark}
\end{figure}

Our starting point is the thermodynamic potential given by,
\begin {align}
 \Omega&= {\cal {U^\prime}}[\Phi,\bar \Phi,T]
    + 2{g_S} {\sum_{f=u,d}} {\sigma_f^2}
    - 6 {\sum_f}{\int_{\lambda}^{\Lambda}}
     \frac {{d^3p}}{{(2\pi)}^3} E_{p_f}\Theta {(\Lambda-{ |\vec p|})}
                                                           \nonumber \\
       &-2{\sum_f}T{\int_\lambda^\infty}\frac {{d^3p}}{{(2\pi)}^3}
     \ln\left[1+3(\Phi+{\bar \Phi}\exp(\frac{-(E_{p_f}-\mu_f)}{T}))
   \exp(\frac{-(E_{p_f}-\mu_f)}{T})+\exp(\frac{-3(E_{p_f}-\mu_f)}{T})
        \right]
                                                            \nonumber\\
       &-2{\sum_f}T{\int_\lambda^\infty}\frac {{d^3p}}{{(2\pi)}^3}
     \ln\left[1+3({\bar \Phi}+{\Phi}\exp(\frac{-(E_{p_f}+\mu_f)}{T}))
   \exp(\frac{-(E_{p_f}+\mu_f)}{T})+\exp(\frac{-3(E_{p_f}+\mu_f)}{T})
        \right]
\end {align}
where $E_{p_f}=\sqrt {p^2+M^2_f}$ is the single quasi-particle energy. In the 
above expression $\sigma_f$ is given as 

\begin {equation}
 \sigma_f=\langle {\bar \psi}_f \psi_f \rangle = - \frac {3{M_f}}{ {\pi}^2} {\int^\Lambda_\lambda}\frac {p^2}{
           \sqrt {p^2+{M_f}^2}}dp,
\end {equation}

We first obtain the mean fields $\sigma$, $\Phi$ and $\bar{\Phi}$ from
the extremization conditions:
$\frac{\partial\Omega}{\partial\sigma} = 0,~~~
\frac{\partial\Omega}{\partial\Phi} = 0,~~~
\frac{\partial\Omega}{\partial \bar{\Phi}} = 0$. 
The field values so obtained are then put back into $\Omega$ to obtain
the thermodynamic potential, which is then used to obtain various
thermodynamic quantities. Some of which have been reported by us in
Ref.~\cite{ab}. For example, the pressure in the finite volume system
is given by,

\begin{equation}
P(T,\mu_q,\mu_I)=-\frac {\partial (\Omega (T,\mu_q,\mu_I)V) }{\partial V}
\label{prs}
\end{equation}
where $T$ is the temperature and $\mu_q$ and $\mu_I$ are the quark and
isospin chemical potentials respectively. The variation of scaled pressure with
$T/T_c$ is shown in figure \ref{pressure}. The critical temperature
$T_c$ is dependent on the system size. We have considered different
system sizes corresponding to $R=2~fm$, $R=2.5~fm$, $R=4~fm$ and
infinite volume. The corresponding values of $T_c$ are 167 MeV, 171 MeV,
183 MeV and 186 MeV respectively. 

\begin{figure}[htb]
\includegraphics[scale=0.45]{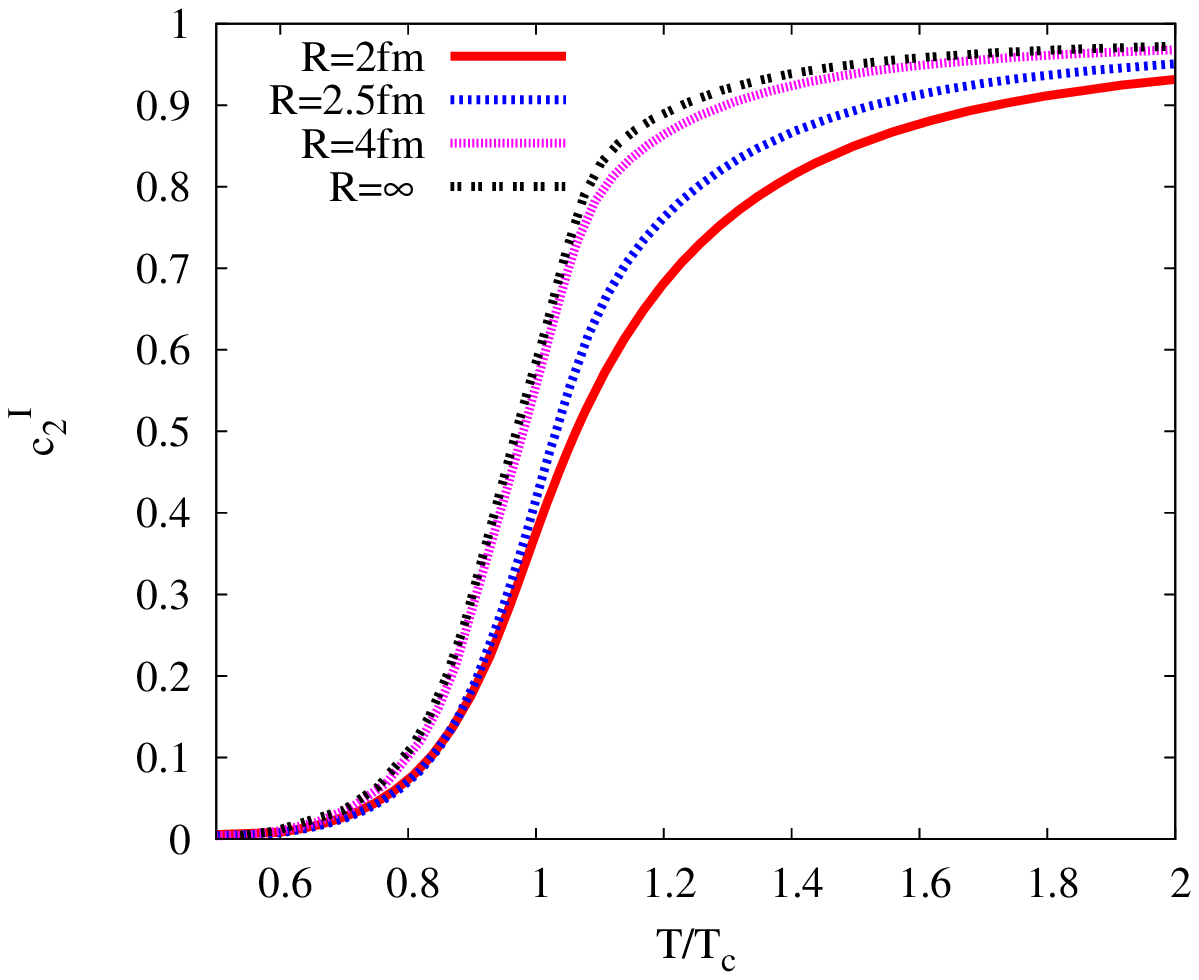}
\includegraphics[scale=0.45]{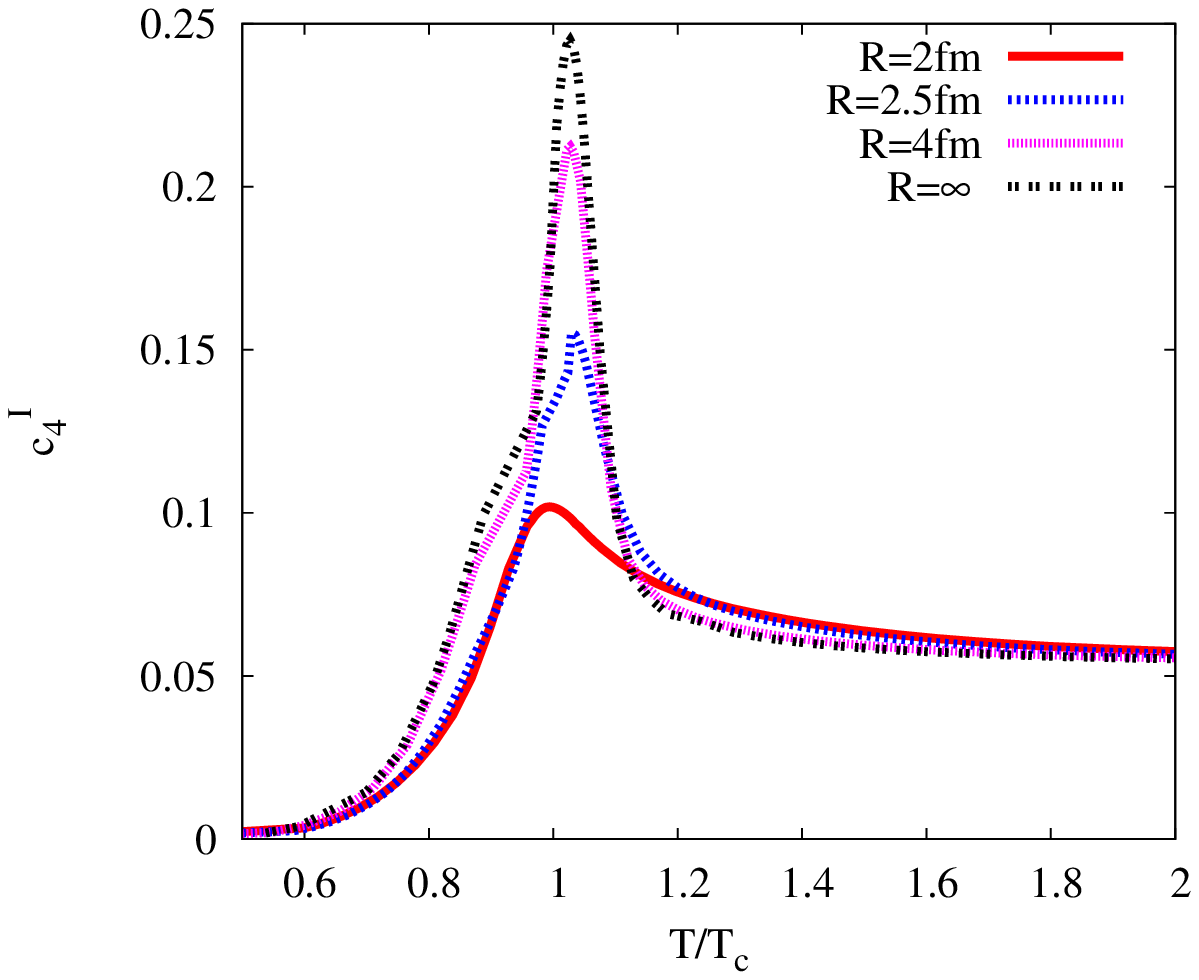}
\includegraphics[scale=0.45]{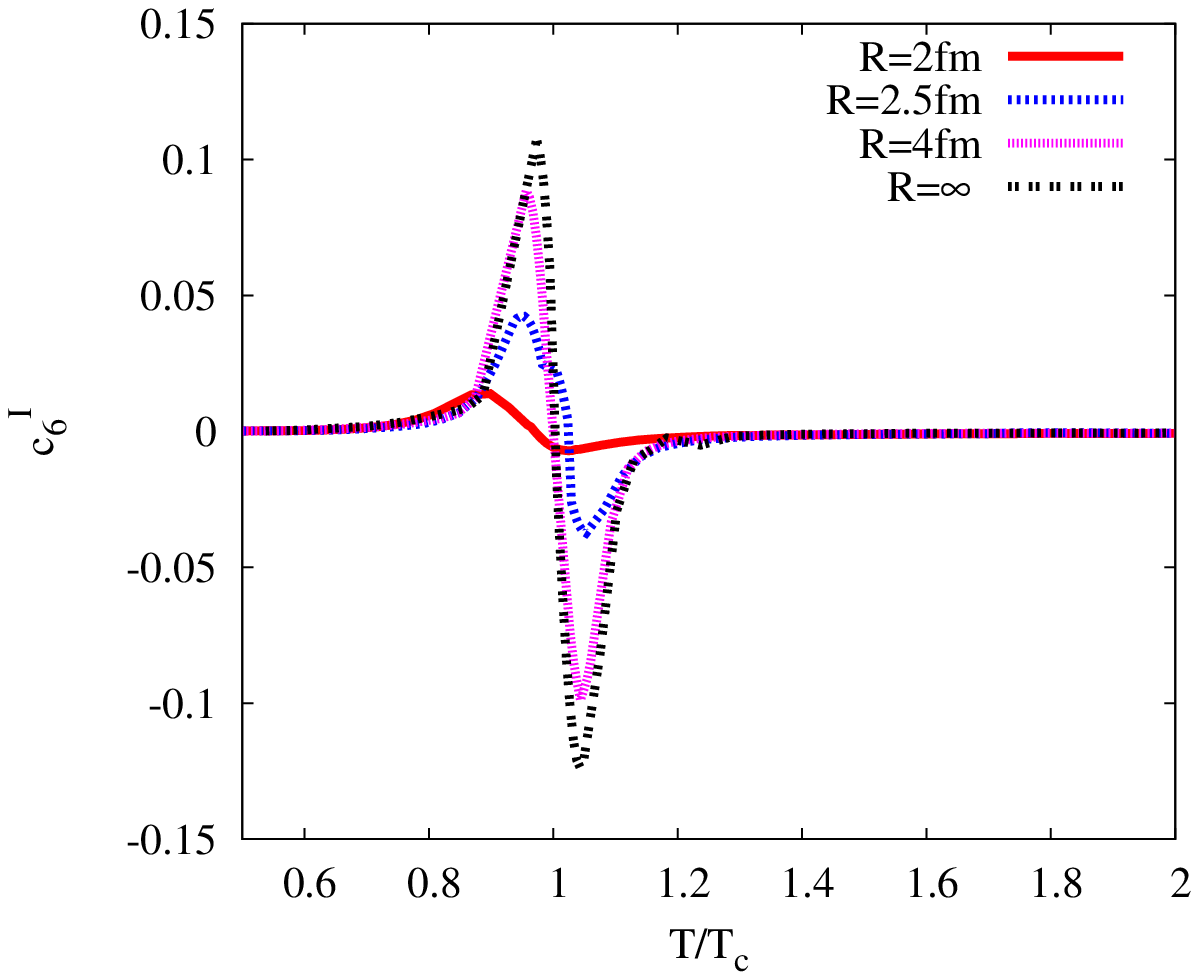}
\caption{(Color online) 
Variation of isospin number susceptibilities with temperature for
different system sizes.}
\label {isospin}
\end{figure}

We now discuss the various susceptibilities of quark number and iso-spin
number. These are defined as,
\begin{eqnarray}
c_n(T) = \frac{1}{n!} \frac
{\partial^n \left ({\Omega(T,\mu_q,\mu_I) / T^4} \right )}
{\partial \left(\frac{\mu_X }{ T }\right)^n}\Big|_{\mu_X=0} ~~~.
\end{eqnarray}
where $\mu_X = \mu_q$ or $\mu_I$. For an expansion around $\mu_X = 0$,
the odd order terms vanish due to CP symmetry. Many of these
susceptibilities have been measured for infinite volume systems in
first principle QCD calculations on the lattice~\cite{Gottieb_prl,
Gottieb_prd,Alton,Gavai_prd1,Gavai_prd2,HOTQCD_08,Cheng_09,WB_12,
HOTQCD_12} as well as hard thermal loop calculations~\cite{Blaizot,
purnendu1,purnendu2,jiang,HTLPT_Najmul,najmul,HMS1_2013,HMS2_2013,
HAMSS_2014,HBAMSS_2014}. At the same time various QCD inspired models
have also made suitable estimates of these fluctuations for infinite
systems (see e.g.~\cite{ray1,ray2,ray3,sasaki_redlich,FLW_2010,abhijit,
abhijit1,roessner1,friman,roessner2,schaefer1,schaefer2,schaefer3,
schaefer4})
Here we present the first computation of finite size effects on these
fluctuations.

\begin{figure}[htb]
\includegraphics[scale=0.45]{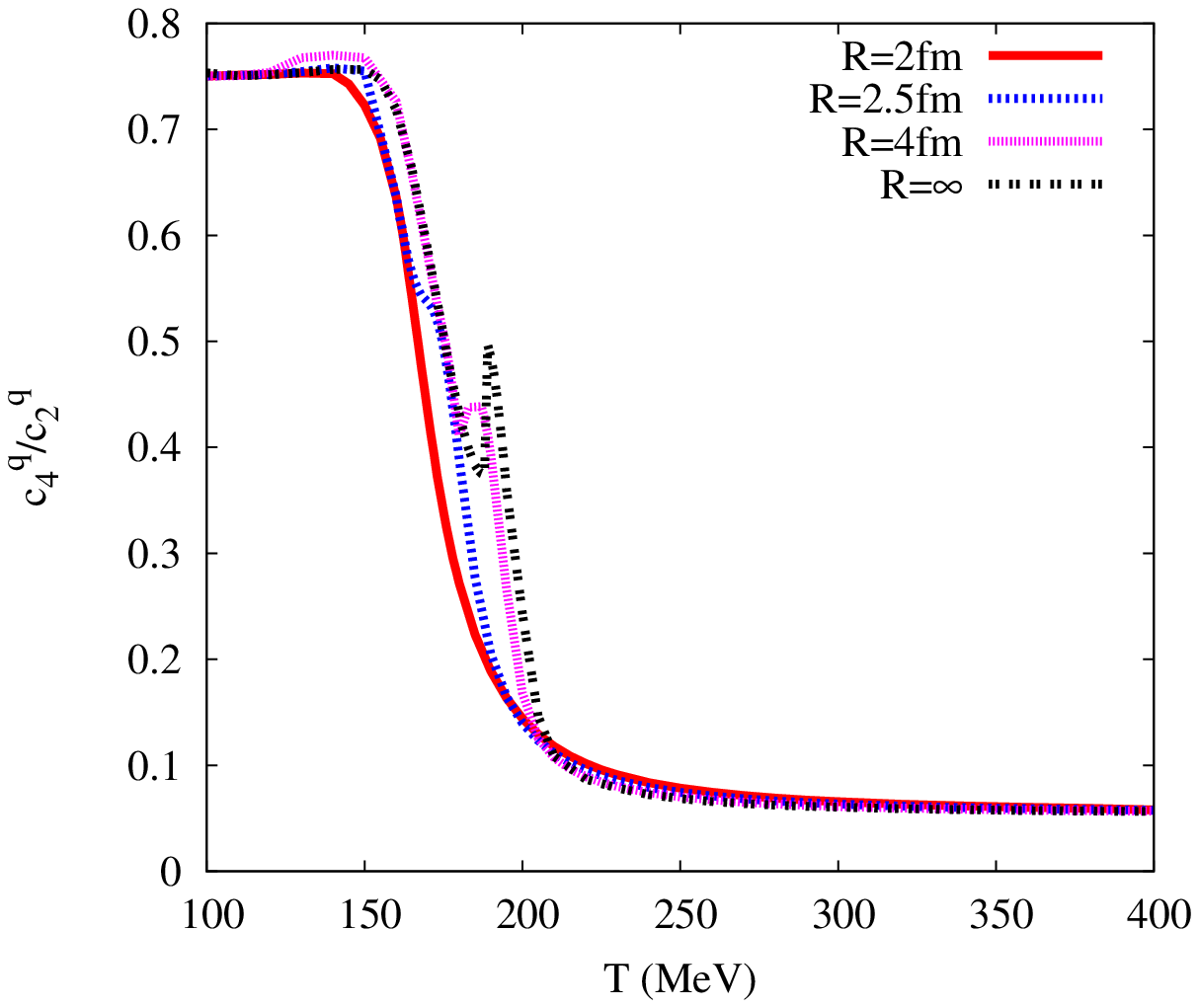}
\includegraphics[scale=0.45]{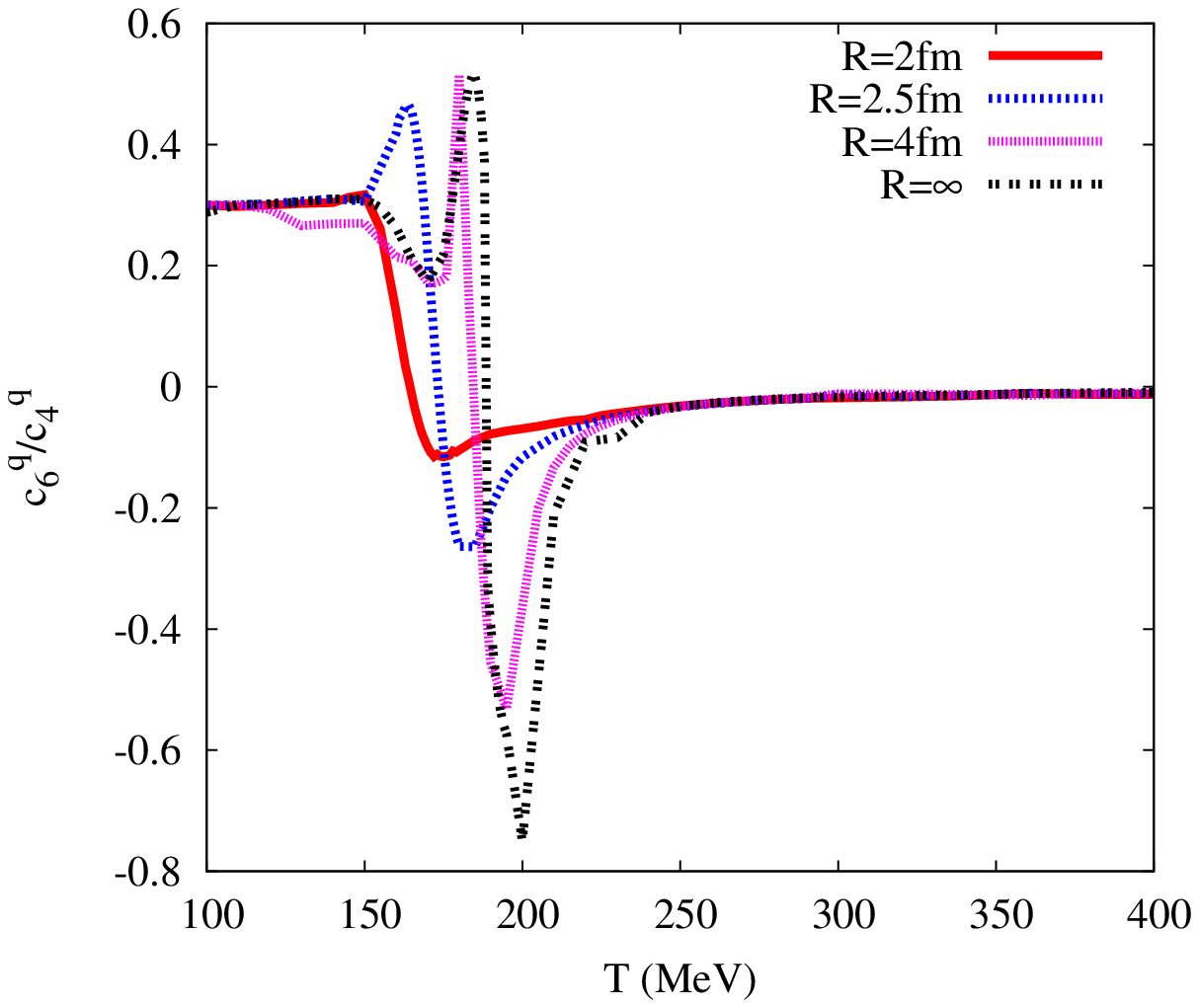} \\
\includegraphics[scale=0.45]{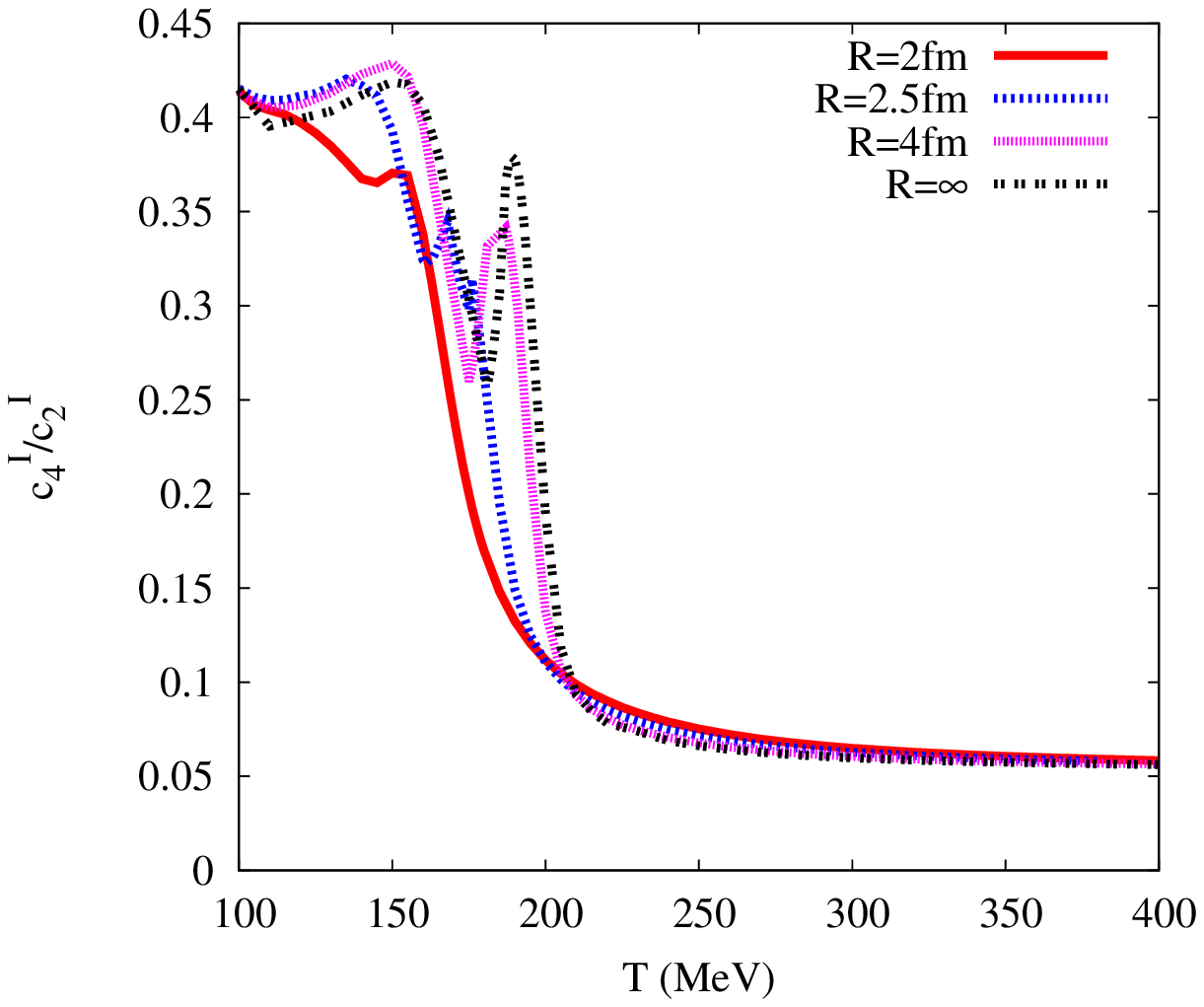}
\includegraphics[scale=0.45]{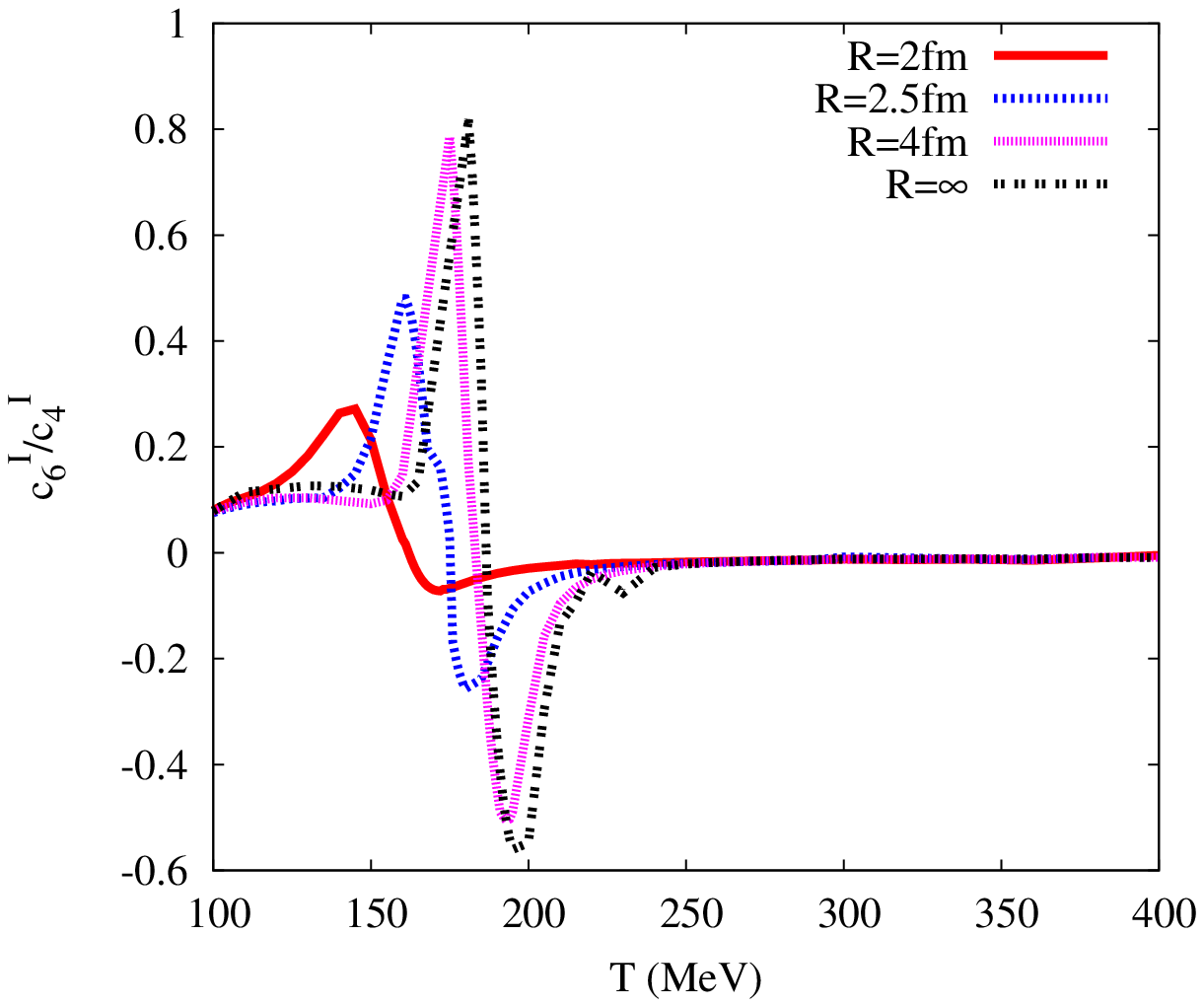}
\caption{(Color online) 
Variation of ratios of fluctuations with temperature for different
system sizes.}
\label {ratios}
\end{figure}

For each system volume considered, we have calculated $\Omega$ at
chemical potentials spaced by 0.1 MeV at a given temperature.
These have been fitted it to an eighth order polynomial in $\mu_X$ using
the GNU plot program. We have chosen the maximum range of $\mu_X$ to be
200 MeV. From the fit we have extracted the coefficients $c_2$, $c_4$
and $c_6$ both for quark number and isospin number susceptibilities.
This procedure has been repeated for different values of temperature.

The variation of quark number susceptibilities with $T/T_c$ are shown
in Fig. (\ref{quark}). The general features for these susceptibilities
in finite volumes are quite similar to that for infinite volume.
However quantitatively we observe significant volume dependence.
With increase in system size there is an enhancement of all the
susceptibilities. For the isospin number susceptibilities shown in
figure \ref{isospin}, we find almost identical behavior. It may be
noted that the most significant finite size effects are seen in the
higher order susceptibilities close to the cross-over region. Given
that the detectors in QGP search experiments are expected to observe
the system frozen close to the cross-over region, one may find an
estimate of the system volume from the measurement of various higher
order fluctuations.

Alternatively, it is important to realize that for a comparison
of fluctuations calculated theoretically with that measured
experimentally, one needs to be confident about the measured system
volumes. Since measuring the system size is quite a difficult task in
the experiments, and one usually resorts to consider ratios of
fluctuations to eliminate the volume factor~\cite{Alton}. However this
assumption is valid when interactions are small and the volume factor
scales out. Therefore in the purely hadronic or partonic phases one
may observe such a scaling of the fluctuations with system size.
However, close to the cross-over region such an assumption may not hold
as large scale fluctuations are dominant and the system deviates from
a stable thermodynamic phase. Now that we have the actual calculations
of system size effects we can easily check the how the ratio of
fluctuations behave. For this purpose we present the ratios
$c_4/c_2$ (kurtosis) and $c_6/c_4$ for both the quark number and isospin
number susceptibilities. In this case we obviously need to plot the
variation with temperature rather than with $T/T_c$. The variations are
shown in Fig. (\ref{ratios}).  We observe that for low and high
temperatures the ratios of fluctuations show the expected scaling with
the system volume, while in the cross-over region there is significant
volume dependence. Thus if the system created in heavy-ion experiments
freeze-out much below $T_c$, the ratios of different susceptibilities
would show the corresponding values for the hadronic phase. The amount
of deviation of these ratios from the hadronic phase results would
indicate the closeness of the system to the cross-over region.

To summarize, we have studied the fluctuations of strongly interacting
matter in a finite volume using the PNJL model. The susceptibilities
in the quark number and isospin number are obtained up to sixth order
for different system sizes. We find a significant volume dependence in
these quantities, which may be useful in analyzing the experimental
data and obtain the size of the fireball formed in the heavy-ion
collision experiments. The volume dependence shows an expected scaling
behavior in the hadronic and partonic phases. In the cross-over region
the system size scaling breaks down and may be use to estimate the
closeness of the created fireball to the cross-over region. Given that
all our present analysis is at zero density the results are suitable
for analyzing LHC data.

The work is funded by Department of Science and Technology (DST) 
(Government of India) and Alexander von Humboldt (AvH) foundation.

\end{document}